\documentclass[format=acmsmall, review=false, screen=true]{acmart}

\usepackage{booktabs} 

\usepackage[ruled]{algorithm2e} 
\renewcommand{\algorithmcfname}{ALGORITHM}
\SetAlFnt{\small}
\SetAlCapFnt{\small}
\SetAlCapNameFnt{\small}
\SetAlCapHSkip{0pt}
\IncMargin{-\parindent}

\usepackage{natbib}
\usepackage{breqn}
\usepackage{bm}
\usepackage{subcaption}
\usepackage{graphicx}
\usepackage[normalem]{ulem}

\begin{document}
\title[]{Cyber Threat Impact Analysis to Air Traffic Flows Through Dynamic Queue Networks}  

\author{Ali Tamimi}
\author{Adam Hahn}
\author{Sandip Roy}
\affiliation{%
  \institution{Washington State University}
  \department{School of Electrical Engineering \& Computer Science}
  \city{Pullman}
  \state{WA}
  \postcode{99163}
  \country{USA}
  }
  \email{ali.tamimi@wsu.edu}
  \email{ahahn@eecs.wsu.edu}
  \email{sroy@eecs.wsu.edu}
\begin{abstract}
Air traffic control  increasingly depends on information and communication technology (ICT) to manage traffic flow through highly congested and increasingly interdependent airspace regions. While these systems are critical to ensuring the efficiency and safety of our airspace, they are also increasingly vulnerable to cyber threats that could potentially lead to reduction in capacity and/or reorganization of traffic flows.  In this paper, we model various cyber threats to air traffic control systems, and analyze how these attacks could impact the flow of aircraft through the airspace. To perform this analysis, we consider a model for wide-area air traffic based on a dynamic queuing network model. Then we introduce three different attacks (Route Denial of Service, Route Selection Tampering, and Sector Denial of Service) to the air traffic control system, and explore how these attacks manipulate the sector flows by evaluating the queue backlogs for each sector's outflows. Furthermore, we then explore graph-level vulnerability metrics to identify the sectors that are most vulnerable to various flow manipulations, and compare them to case-study simulations of the various attacks. The results suggest that Route Denial of Service attacks have a significant impact on the target sector and lead to the largest degradation to the overall air traffic flows. Furthermore, the  impact of Sector Denial of Service attack impacts are primarily confined to the target sector, while the Route Selection Tampering impacts are mostly confined to certain aircraft.

 
\end{abstract}

%
%
%
%


\maketitle


\section{Introduction}

Air Traffic Control (ATC) systems are responsible for ensuring the safety and efficiency of airspace. The primary function of ATC systems  is to guide each aircraft from the departure gate to the arrival gate along planned routes, in such a way as to avoid conflicts. At longer time horizons, ATC systems are also responsible for scheduling and routing aircraft to match demand with capacity, in a way that is efficient for the stakeholders -- this
additional function is known as {\em air traffic management}. Traffic control requires continuous communication between each aircraft and a number of ATC facilities (towers, control centers) during different phases of the flight.  In addition, for both traffic control and management, various facilities (including towers and control centers, as well as airline dispatch offices and central command elements) must communicate to decide aircraft flight plans, enact modifications, and exchange information about traffic. 

While air traffic control has historically been dependent on radar and VHF-based voice communication, modern air traffic control systems are increasingly using digital technology to enhance the control and awareness of the airspace. For example, the U.S. Federal Aviation Administration (FAA) is working on the transition to the Next Generation Air Transportation System (NextGen). NextGen intends to modernize air traffic control to increase the capacity and reliability of the airspace, while also improving safety and security, and minimizing the environmental impact of aviation \cite{NextGen}. NextGen expands the use of digital communication through the Automatic Dependent Surveillance-Broadcast (ADS-B) system which requires that aircraft broadcast their location, rather than depending on radars. In addition, NextGen envisions a host of new decision-support tools for traffic control and management, which will require automated communication as well as ingestion of data sources (e.g. ensemble weather forecasts) from public-domain websites. While these advances should ideally lead to improved control and fewer delays, they also expand the system's attack surface and expose the system to threats of digital manipulation. 

While NextGen technologies increase data sharing throughout the NAS to improve system operations, this also introduces additional cybersecurity challenges~\cite{mitrenas}. The threat of cyber attack to air traffic control has been well documented in both government reports and in academic literature.  FAA reports  identify the need to improve the cybersecurity and resilience of the National Airspace (NAS)~\cite{faanas}, while the National Academy of Sciences has recommended the development of improved threat models to explore the risk of air transportation system architectures ~\cite{nasnas}.  While the U.S. Government Accountability Office (GAO) has released numerous reports identifying potential vulnerabilities in these systems \cite{wilshusen2015information,dillingham2015air}. There have also been real-world intrusions to air traffic control, as reports suggest attacks targeting British Air Traffic Control attempted to manipulate the voice (VHF) communications \cite{HackAirTraffic}, while sophisticated cyber attacks also infiltrated Sweden's ATC \cite{swedenHack}. Beyond cyber attacks, a number of high-profile failures as well as a physical-world attack have impacted the cyber infrastructure associated with the air traffic control system.

The potential impacts of cyber attacks and failure to air traffic control are multifacted, as the recent events have demonstrated.  First, such events may directly impact system safety, by interfering with controllers' ability to guide aircraft.  In addition, however, these events have cascading impacts on the wide-area management of traffic, thus complicating scheduling and routing, reducing system efficiency, increasing controller workload, and potentially indirectly degrading safety at remote locations.  Safeguards are in place to reduce the risk of direct safety impacts, but the wide-area impacts of cyber attacks on traffic control and management are not well understood, and are a crucial concern.  While current academic literature has explored air traffic control system security and vulnerabilities, most of this work has focused on identifying technical vulnerabilities in air traffic control protocols, but has not yet explored how such vulnerabilities impact system-level aircraft routing and flows. On the other hand, models for system-level air traffic flows and their management have been developed, but none of these studies realistically model the cyber system and associated attack surface.  This work aims to bridge the gap, by modeling the impact of cyber attacks on air traffic flows, and analyzing attack impacts on regional air traffic management. To perform this analysis, we present a model for air traffic flows management based on the dynamic queuing network introduced in \cite{wan2013dynamic}. We then model various attacks from previous literature within this network and calculate the impacts of these various attacks on air traffic flows. Moreover, we apply the metric presented in \cite{royvulnerability} to identify vulnerable routes and sectors in air traffic flows.

The remainder of the paper is organized as follows,  Section 2 introduces  related work, Section 3 presents the air traffic control system and dynamic queuing network model, Section 4 present our threat model and flow manipulations, Section 5 provides system-level risk analysis from ATC attacks, and Section 6 provides discussion on the resulting risk. 

\section{Related Works} 
There have been many previous efforts exploring both the cybersecurity vulnerabilities within ATC system, along with work exploring the vulnerability of ATC sectors and flows to traditional disruptions (e.g., weather). Key efforts that explored the cybersecurity of air traffic control include work by Strohmeier \cite{strohmeiersecurity, Strohmeier2016} which introduces a survey of the communication devices used in ATC and identifies vulnerabilities for each device. Further work in \cite{costin2012ghost}, explores cyber vulnerabilities within the NEXTGEN platform, specifically the ADS-B communications and then explores potential threat impacts based on the aircraft locations and attacker goals. In other work, they simulate an environment including an air traffic model, existing surveillance systems, ADS-B systems, and wireless channel model to investigate communication infrastructure of next generation air traffic management and evaluate the performance of communication and optimize it \cite{park2012investigating,park2015performance}. A variety of other works have also explored vulnerabilities within ADS-B. In \cite{schafer2013experimental}, they evaluated ADS-B attacks and quantified various factors, such as the attacker's location and the signal strength, based on their ability to manipulate various aircraft messages. In \cite{purton2010identification}, the authors explore different potential attacks against ADS-B systems, including network intrusions, message spoofing, and communication malfunction, and analyzed each based on their threats, attack opportunities, weaknesses, and strengths. Broader ADS-B attack taxonomies and potential impacts are identified in \cite{McCallie201178}, including the techniques required for exploitation and their difficulty. A model-based approach that investigates various physical and cyber vulnerabilities of ADS-B from different aspects is presented in \cite{thudimilla2017multiple}.

Other related work has explored theoretic models and analysis techniques to evaluate how interruptions within air traffic sectors impact the broader system-level flows, i.e. the flow management function of the air traffic control system.  In \cite{gwiggner2b10}, the authors reviewed recent models of air traffic flow analysis and then explored queuing networks, traffic flow theory, and cellular automata methods to discover relationships between system variables. They conclude that the combination of model-based flow analysis and analysis of flight data leads to new insight into the air traffic congestion mechanisms. In  \cite{sridhar2008modeling}, the authors categorized the traffic flow models into three groups, including linear dynamic system models, other Eulerian models, and partial differential equation models, which can enable simplified analysis of wide-area dynamics. Furthermore, in \cite{roysridhar,bayen2004eulerian,menon2006computer} the authors explored Eulerian network models for air traffic flows.
These flow-level models for air traffic were later enhanced to represent traffic at varied resolutions, to explicitly capture multiple origin-destination pairs, and to model management initiatives as queueing elements \cite{mitre1,mitre2}.
Finally, agent-based modeling is explored in \cite{wang2009analysis} where each flight and control agent is defined as an agent and used to analyze different system properties such as throughput, capacity, delay, delay jitter, and congestion. Furthermore, domain-specific multi-agent system models have also been explored in the domain of air traffic management to provide constructs (which can be instantiated to implement specific tasks and procedures in air traffic management domain) for different scenarios \cite{rungta2016modeling}.  Building on these studies, a series of recent works have begun to assess the propagative impacts of traffic flow restrictions -- whether due to cyber events, weather, or other causes (e.g. space vehicle operations) \cite{royvulnerability,royvuln2,royvuln3}.

 The main purpose of this article is to pursue comprehensive modeling and analysis of the impacts of cyber- attacks on regional air traffic flows.  Toward this goal, we pursue integrated modeling of classes of cyber attacks and queueing-type models of wide-area traffic evolution, and explore simulation-based as well as graph-theoretic approaches to impact analysis. Specific contributions include:

\begin{itemize}
\item Categorization and modeling of threats to communications that can impact en-route control of air traffic.
\item Integration of these threat/attack models with queueing-theoretic models for air traffic flows.
\item Threat analysis of the proposed attacks using simulations of the integrated model.
\item Simplified vulnerability analysis of the air traffic system using algebraic graph theory approaches, and comparison with simulation results.
\end{itemize}
Throughout the development, case studies are presented to illustrate and evaluate the methodology.

\section{Air Traffic Control System and Threat Model}

\begin{figure*}
\centering
\begin{minipage}{.3\textwidth}
\centering
\includegraphics[width=\linewidth]{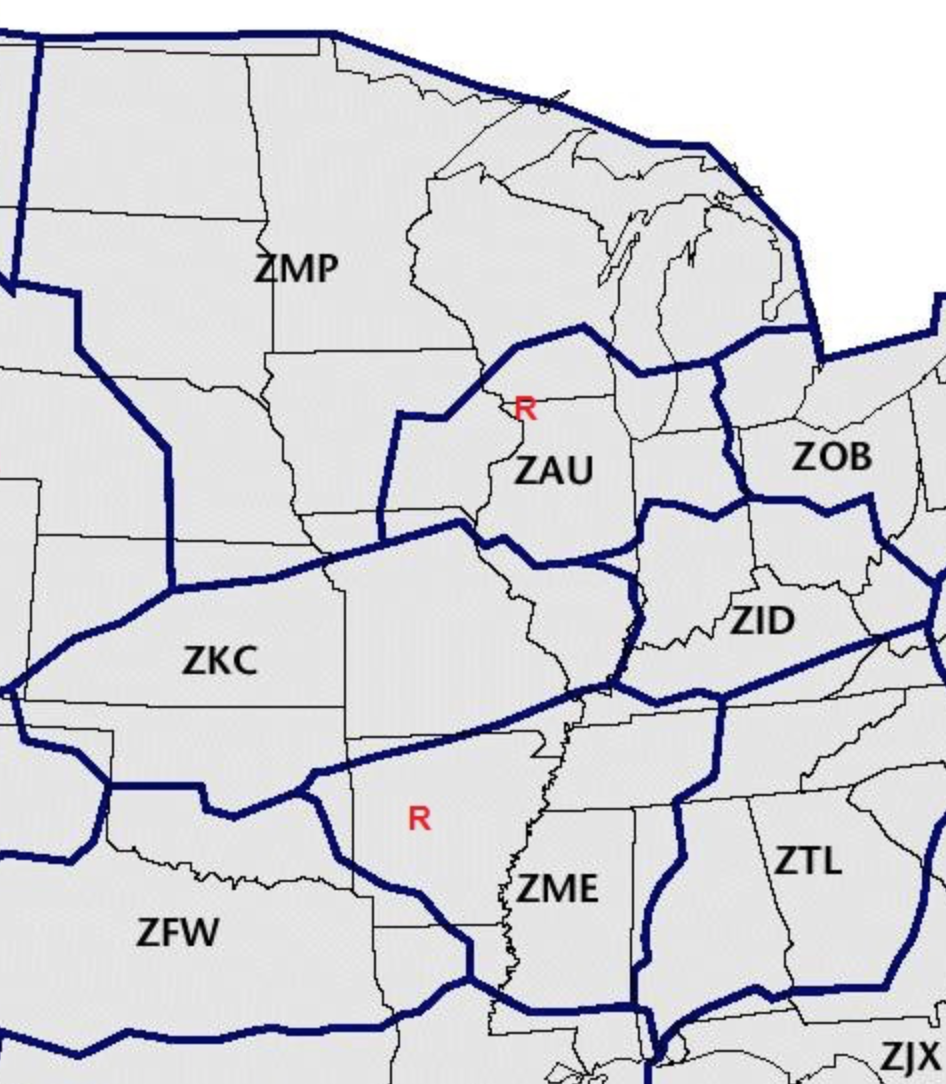}
\caption{A part of the map of Air Route Traffic Control Centers}
\label{fig:map}
\end{minipage}\hfill
\begin{minipage}{.3\textwidth}
\centering
\includegraphics[width=\linewidth]{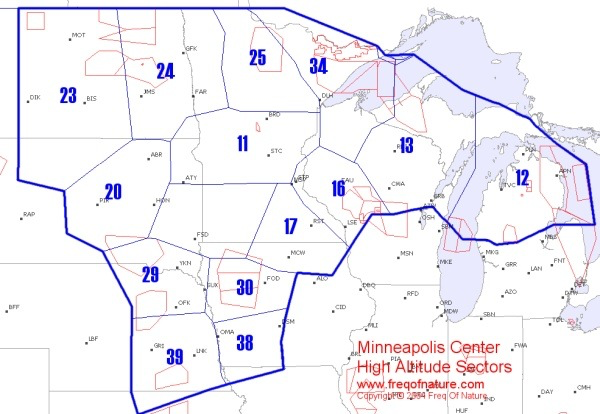}
\caption{Sectors of Minneapolis ARTCC (ZMP)}
\label{fig:zmpsec}
\end{minipage}\hfill
\begin{minipage}{.3\textwidth}
\centering
\includegraphics[width=\linewidth]{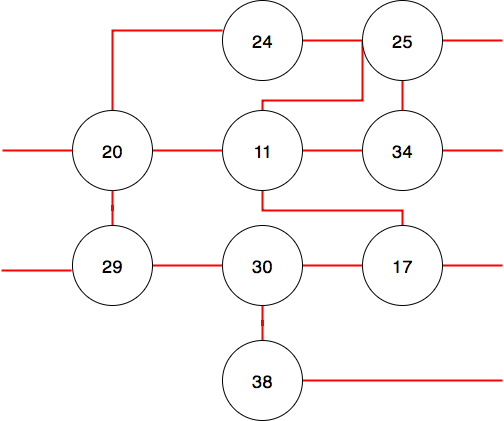}
\caption{Graph model of ZMP ARTCC which is shown in Fig. \ref{fig:zmpsec}}
\label{fig:graph}
\end{minipage}
\end{figure*}

This section will provide an introduction to Air Route Traffic Control Center (ARTCC) operations and the aircraft communications required to support it. ARTCC is required to support each phase of a flight including: preflight, takeoff, departure, en-route, descent, approach, and landing \cite{faa2015}. The main function of ARTCCs is to ensure appropriate aircraft separation and the safety of the various sector routes \cite{nolan2010fundamentals}. ARTCCs accept aircrafts and pass them to other ARTCCs or terminal control centers. There are 22 ARTCCs which are located in nineteen states \cite{ARTCC2013}. Fig. \ref{fig:map} shows a subsection of the U.S ARTCC, where each ARTCC is giving a unique name (e.g., ZMP refers to Minneapolis ARTCC). Furthermore, each ARTCC consists of several sectors as demonstrated in Fig. \ref{fig:zmpsec}, which shows that sectors of the ZMP ARTCC. This work will utilize a graph model to model and analyze the flow of these sectors, as demonstrated in Fig. \ref{fig:graph} which provides the graph for the ZMP ARTCC.

\begin{figure*}[h!]
\centering
\includegraphics[width=\textwidth]{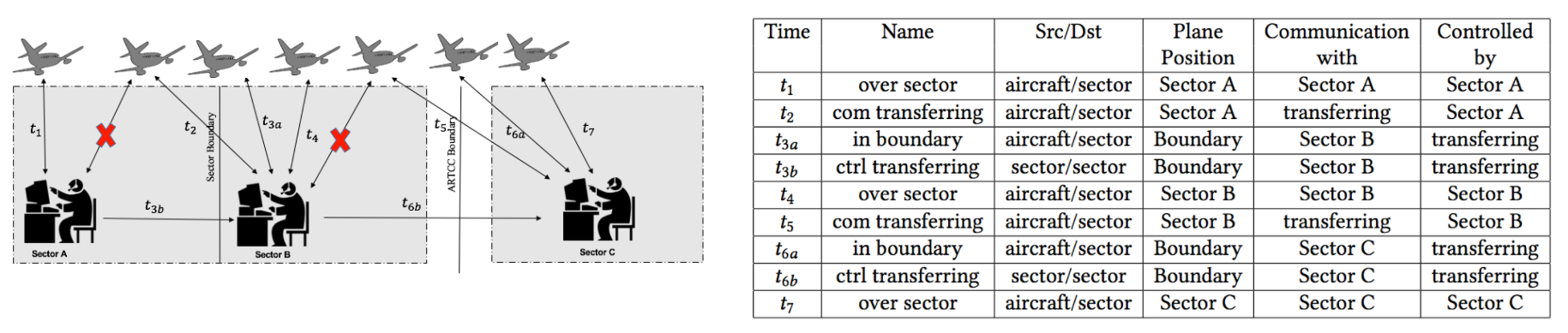}
\caption{Communications in the en-route phase when aircraft routes between the sectors and a table which shows when and where each communication is established }
\label{fig:enroutephase}
\end{figure*}

This section will provide an overview of how  ARTCCs manage the flow of aircraft through sectors. When an aircraft enters a sector, it communicates to the corresponding sector controller. Each sector is controlled by one or a team of controllers and is responsible for separating of the aircrafts. At the sector boundary, they are responsible for transferring control from the previous controller. The process of transferring control and transferring communication is called the ``hand-off''. Fig. \ref{fig:enroutephase} shows the communications of en-route phase.
The explanation of each time interval is as follow:
\begin{itemize}
\item \textit{\textbf{Over Sector:}}\ The aircraft flies over the sector and is controlled by that current sector, its only communication is with the current sector in this phase.
\item \textit{\textbf{Com transfer:}}\ The aircraft is close to the sector boundary and it is still controlled by current sector, but the communication will begin its transfer to the next sector before it reaches the boundary.
\item \textit{\textbf{In boundary:}}\ The aircraft flies over the boundary and keeps the communication with next sector.
\item \textit{\textbf{Ctrl transferring:}}\ The aircraft flies over the boundary and the control is passed from the current sector to the next sector.
\end{itemize}
Based on the above explanation, Fig. \ref{fig:enroutephase} shows the communication of an aircraft during the en-route phase as it passes to different sectors and ARTCCs. It also provides a table that identifies the location between sectors, its communication to sector controllers and the controlling sector. At $t_1$, the aircraft communicates with sector A, which is responsible for control the aircraft. Before leaving the sector A ($t_2$), the communication is transferred to Sector B. Although in this step the communication is transferred to sector B, the aircraft is still controlled by sector A. $t_3$ demonstrates the boundary of two sectors, Sector A passes the control of aircraft to the Sector B. $t_4$, $t_5$, and $t_6$ are same as $t_1$, $t_2$, and $t_3$ respectively, except that at $t_5$ and $t_6$ the communication and control are passed to the sector in another ARTCC.

\subsection{En-route Communications and Threats}
\begin{figure}[h!]
\centering
\includegraphics[width=3.5 in,height=3 in]{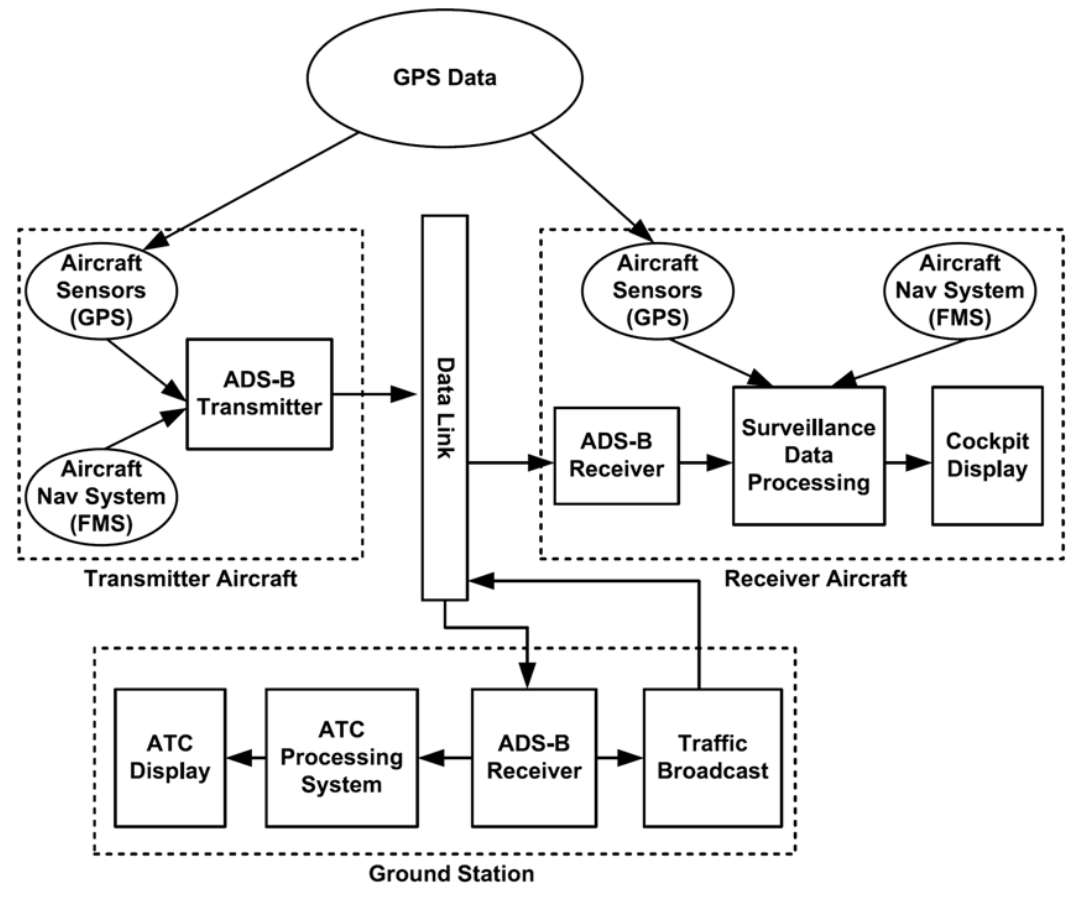}
\caption{ADS-B System Architecture\cite{schafer2013experimental} }
\label{fig:ADSB}
\end{figure}

Automatic Dependent Surveillance-Broadcast (ADS-B) is the foundation of the NextGen air traffic control infrastructure and will become mandatory in the U.S. by 2020 \cite{schafer2013experimental,costin2012ghost}. The main function of ADS-B is to improve aircraft surveillance by having planes broadcast their altitude, airspeed, and location to other aircrafts and ground stations. Fig. \ref{fig:ADSB} shows the architecture of ADS-B system \cite{schafer2013experimental}. ADS-B uses GPS to receive location data and then advertises the aircraft's position and velocity through a 1090 MHZ data link.

\indent A core problem of ADS-B is the lack of security, specifically message encryption and authentication. Recent research has identified how attackers could manipulate these vulnerabilities to manipulate the ARTCC operations \cite{costin2012ghost} through the following attacks:
\begin{itemize}
\item\textit{\textbf{Ghost Injection:}}\ In this attack, the attacker broadcasts ADS-B messages for a non-existing aircraft (ghost aircraft). This results in the non-existing aircraft appearing on the radar screen of the ground station with other valid  aircrafts. During such an attack, it may not be possible for the ground station to distinguish between the real aircrafts and non-existing aircrafts. This confusion may lead to a station denial of service attack as the controllers can no longer correctly serve the existing aircraft.
\item\textit{\textbf{Flood Denial:}}\ In this attack, the attacker produces a jamming signal that disrupts the 1090 MHz data link. As a result, the ground station cannot receive aircrafts' ADS-B messages and the aircraft may disappear from the controller screen of ground station. Therefore, the ground station is unable to send a controlling message to the aircrafts and they will continue their routes regardless of any command.
\item\textit{\textbf{Virtual Trajectory Modification:}}\ By selectively jamming an aircraft's message and replacing it with a modified message, an attacker can report the wrong location of the aircraft. Therefore, the ground station will have non-existing aircraft on its screen instead of the real aircraft. As a result of this attack, the controller manages the flows based on the wrong assumption which is the non-existing aircraft.
\end{itemize}

Based on these previously introduced threats, we propose three attack scenarios to identify how each threat manifests into a impact on ARTCC flows. In the next section, we explain how these attacks influence air traffic flows using the DQN model. 

\textbf{Scenario 1: Route Denial of Service (RDOS):} An attacker injects spoofed aircrafts in a specific route using the Ghost Injection attack. Therefore, the air traffic controller sees a combination of real and ghost aircraft on their displays. Since the controller is unable to distinguish between real and ghost aircraft, it cannot serve the real aircrafts and will shut down the targeted route to prevent any spacing violations. Therefore, the aircrafts cannot be served and have to stay in their current sectors. If the attacker is only targeting one route, it is called a partial RDOS (P-RDOS). If the attacker is targeting all of the outflows of a sector, it is called a complete RDOS (C-RDOS).

\textbf{Scenario 2: Route Selection Tampering (RST):} In the second scenario, the attacker modifies the ADS-B message of an aircraft using the Virtual Trajectory Modification attack. Therefore, the route of aircraft is changed on the controller screen. Therefore, the controller manages a flow including a non-existing aircraft. On the other hand, since the messages of the real aircraft are jammed, it is not included in the flow management.

\textbf{Scenario 3: Sector Denial of service (SDOS):} In this scenario, the attacker applies a Flood Denial attack to a sector. The attacker sends the jamming signal to the VHF channel of a sector controller and the controller cannot send any instruction related to queue management to the aircrafts. Therefore, the aircrafts continue their routes without queuing management. This attack increases traffic of the outgoing routes. Therefore, the next sectors face the unpredicted traffics and the backlog of their queues increases which leads to delay.

\subsection{DQN Model for Air Traffic Flows Management}
\begin{figure}[h!]
\centering
\includegraphics[width=2 in,height=1 in]{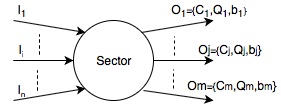}
\caption{The DQN model parameters }
\label{fig:nodemodel}
\end{figure}

In this section, we present an air traffic flow model based on a dynamic queuing network (DQN). In the graph model, each node presents a sector where the inflows and outflows are represented by directed edges. Therefore, if there is a route from sector A to sector B, there will be a directed edge from node $A$ to node $B$. Fig. \ref{fig:nodemodel} shows the graph model of a sector. $I_{i}$ is an inflow which comes from another sector. Each aircraft that comes through $I_{i}$ should leave the sector through one of the outflows ($O_{i}$) based on its flight route. Each outflow has three parameters. $C$ is the capacity of outflow and shows the number of aircrafts that can be served by the sector through the outflow in each time interval. $Q$ is the queue of outflow and contains the aircrafts that are waiting to be served. The number of aircraft waiting in the queue in each time interval is the backlog which is presented by $b$. In other words, $b$ is the number of elements of $Q$.

In our model, The state variable is $X_{i}(t)$ that is the backlog of the $i$-th outflow at time $t$. The number of aircrafts which comes into the queue of outflows is represented by $U_{i}(t)$, while the number of aircrafts which be served is smaller or equal to the capacity of outflow ($C$). Therefore,  an attacks can be modeled as a changes the flow graph structure through edge removals or manipulations to edge flows. The model also defines $U_{j}$ as the number of aircraft which come to the queue of output $O_{j}$. $I\xrightarrow[\text{}]{routes}U$ is the mapping function which finds the corresponding outflow for each aircraft based on the flight route. Our model is as follow:

\begin{equation}
\begin{bmatrix}
    x_{1}(t) \\
    \vdots \\
    x_{n}(t)
\end{bmatrix}
=
\begin{bmatrix}
    U_{1}(t) \\
    \vdots \\
    U_{n}(t)
\end{bmatrix}
+
\begin{bmatrix}
    x_{1}(t-1) \\
    \vdots \\
    x_{n}(t-1)
\end{bmatrix}
-
\begin{bmatrix}
    C_{1} \\
    \vdots \\
    C_{n}
\end{bmatrix}
\label{eq:model1} 
\end{equation}
\begin{itemize}

\item $x_i(t)$ is the backlog of the sector for the outflow number $i$ at time $t$,
\item $I_i(t)$ is the number of airplanes that come to the sector through the inflow number $i$ at time $t$,
\item $C_i $ is the number of airplanes that can be served at the outflow number $i$ of the sector in each time interval,
\item $U_i(t) $ is the number of airplanes leaving the sector through the outflow number $i$ in each time interval.
\end{itemize}

 If the value of $x_i(t)$ is negative, the number of aircrafts in the queue (backlog of last time interval) plus the number of aircrafts that come into the queue in the current time interval is smaller than the $i$th outflow capacity and sector controller can serve all of the aircrafts of the $i$th outflow . In this condition ($x_i(t) < 0$), we consider $x_i(t)=0$.

\section{Threat Analysis}

In this section, we demonstrate the  attack scenarios introduced in Section 3.1 on the proposed DQN model to enable the analysis of their impact to air traffic flows. 

\subsection {Route Denial of Service (RDOS)}
\begin{figure}
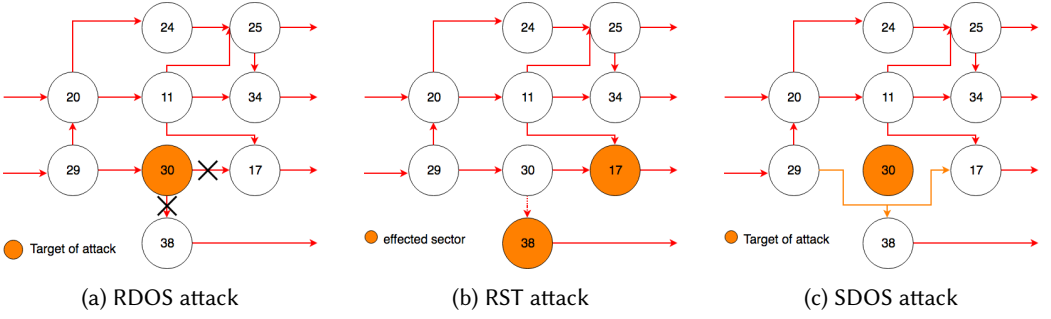

\begin{subfigure}{0.31\textwidth}
\includegraphics[width=\linewidth]{figures/RDOSAttackBig.png}
\caption{RDOS attack} \label{fig:map2}
\end{subfigure}
\hspace*{\fill} 
\begin{subfigure}{0.31\textwidth}
\includegraphics[width=\linewidth]{figures/RSTAttackBig.png}
\caption{RST attack} \label{fig:map5}
\end{subfigure}
\hspace*{\fill} 
\begin{subfigure}{0.31\textwidth}
\includegraphics[width=\linewidth]{figures/SDOSAttackBig.png}
\caption{SDOS attack} \label{fig:map4}
\end{subfigure}
\caption{Attacks demonstration in flow graph} \label{fig:attackdemonstration}
\end{figure}

The RDOS attack assumes that the attacker may cause a controller screen change by injecting non-existing aircrafts. Since the controller is unable to distinguish between the real and non-existing aircrafts, it leads to a route or sector shutdown. If there is a route shutdown (P-RDOS), the controller cannot serve the aircrafts through the targeted route and should reroute the aircrafts in the queue. If there is a sector shutdown (C-RDOS), other sectors should reroute the aircrafts which are heading to the targeted sector and the aircrafts which are already in the queue of targeted sectors should wait until the attack is recognized. We define the attack model for C-RDOS  as follow.

\begin{equation}
\begin{bmatrix}
    x_{1}(t) \\
    \vdots \\
    x_{n}(t)
\end{bmatrix}
=
\begin{bmatrix}
    u_{1}(t) \\
    \vdots \\
    u_{n}(t)
\end{bmatrix}
+
\begin{bmatrix}
    x_{1}(t-1) \\
    \vdots \\
    x_{n}(t-1)
\end{bmatrix}
-
(1-b(t))\times
\begin{bmatrix}
    C_{1} \\
    \vdots \\
    C_{n}
\end{bmatrix}
\label{eq:model2} 
\end{equation}
The attack model for P-RDOS is defined as follow.
\begin{equation}
\begin{bmatrix}
    x_{1}(t) \\
    \vdots \\
    x_{n}(t)
\end{bmatrix}
=
\begin{bmatrix}
    u_{1}(t) \\
    \vdots \\
    u_{n}(t)
\end{bmatrix}
+
\begin{bmatrix}
    x_{1}(t-1) \\
    \vdots \\
    x_{n}(t-1)
\end{bmatrix}
-
\begin{bmatrix}
   1 - b_{1}(t) \\
    \vdots \\
   1 - b_{n}(t)
\end{bmatrix}\cdot
\begin{bmatrix}
    C_{1} \\
    \vdots \\
    C_{n}
\end{bmatrix}
\label{eq:model2} 
\end{equation}

\indent where $b$ is defined as follow: 
\[   
b(t) = 
     \begin{cases}
       \text{1,} &\quad\text{complete attack happens}\\
       \text{0,} &\quad\text{no attack}\\
            \end{cases}
\]
\[   
b_{i}(t) = 
     \begin{cases}
       \text{1,} &\quad\text{partial attack happens at route number \textit{$i$}}\\
       \text{0,} &\quad\text{no attack at route number \textit{$i$}}\\
            \end{cases}
\]
\indent In the graph model, the outflow edges that correspond to the targeted sector should be removed. Suppose that in Fig. \ref{fig:graph}, sector 30 is the target of attack. Fig. \ref{fig:map2} shows the structure of graph after the RDOS attack. 

\subsection {Route Selection Tampering (RST)}
In a RST attack, an attacker could jam the ADS-B signal of an aircraft and inject new ADS-B messages of non-existing aircraft using Virtual Trajectory Modification attack. In this condition, aircraft maintains the original route, however, there is a route change on the controller screen that shows the aircraft is heading to another sector. In this attack, the sector which is the destination of non-existing aircraft has an additional aircraft (non-existing aircraft) in the queue management. In this attack, we show vector $b$ as follow.
\[   
b_i = 
     \begin{cases}
       \text{1,} &\quad\text{attack effects on outflow number $i$}\\
       \text{0,} &\quad\text{no attack}\\
            \end{cases}
\]
\indent If more than one aircraft is targeted, the value of $b$ can be greater than one.
\begin{equation}
\begin{bmatrix}
    x_{1}(t) \\
    \vdots \\
    x_{n}(t)
\end{bmatrix}
=
\begin{bmatrix}
    u_{1}(t) \\
    \vdots \\
    u_{n}(t)
\end{bmatrix}
+
\begin{bmatrix}
    b_{1}(t) \\
    \vdots \\
    b_{n}(t)
\end{bmatrix}
+
\begin{bmatrix}
    x_{1}(t-1) \\
    \vdots \\
    x_{n}(t-1)
\end{bmatrix}
-
\begin{bmatrix}
    C_{1} \\
    \vdots \\
    C_{n}
\end{bmatrix}
\label{eq:model3} 
\end{equation}

\indent Fig. \ref{fig:map5} shows the RST attack. After leaving sector 30, the attacker jams the ADS-B messages which are received by sector 17 and inject the new aircraft to the route which heading to sector 38. Both sectors 17 and 38 are affected by the attack. Although sector 17 cannot receive ADS-B messages of aircraft, the real aircraft is heading to this sector and should be served. Moreover, a non-existing aircraft is shown on the controller screen of Sector 38 and should be managed by this sector.

\subsection {Sector Denial of Service (SDOS)}
In this section, we introduce the SDOS attack and present our model. In the SDOS attack, after establishing the voice communication between the aircraft and the receiving sector, the attacker jams this communication. Since the sector cannot send messages to the aircraft, the aircraft continues its route without queue management. As a result of the SDOS attack, the sector cannot inform the aircrafts about the over service capacity condition. In other words, all the aircrafts in the sector continue their routes without waiting in the queues. The consequence of this attack is more visible in the sectors that are after the targeted sector. We present the model of this attack as follow. Since the aircrafts do not come into the queues, the backlog is equal to zero.
\begin{equation}
\begin{bmatrix}
    x_{1}(t) \\
    \vdots \\
    x_{n}(t)
\end{bmatrix}
= 0
\label{eq:model2} 
\end{equation}

Since the attack hinders queue management, the outflows of the targeted sector increases with no limitation and changes the inflows of the sectors that are after it. Therefore,  aircraft will ignore the queues of the targeted sector and continue their routes. In the flow graph model, the node which represents the targeted sector is removed and the outflow edges of previous sectors connect to the inflow edges of next sectors. Suppose that in Fig. \ref{fig:graph}, Sector 30  is the target of SDOS attack. Fig. \ref{fig:map4} shows the changes in the graph structure after the attack. The major impact of this attack is in the sectors that have an inflow from the targeted sector. In Fig. \ref{fig:map4} the largest impact is found in sectors 17 and  38. The number of aircrafts  in their queues increases and it needs more time to serve them. The reason is that there is no queue management in sector 30 and all aircrafts are directed to next sectors without staying in the queues. As a result of this situation, the number of aircrafts increases in the queues of sector 17 and sector 38.

\begin{figure}[h!]
\centering
\begin{minipage}{.5\textwidth}
  \centering
  \includegraphics[width=1\linewidth]{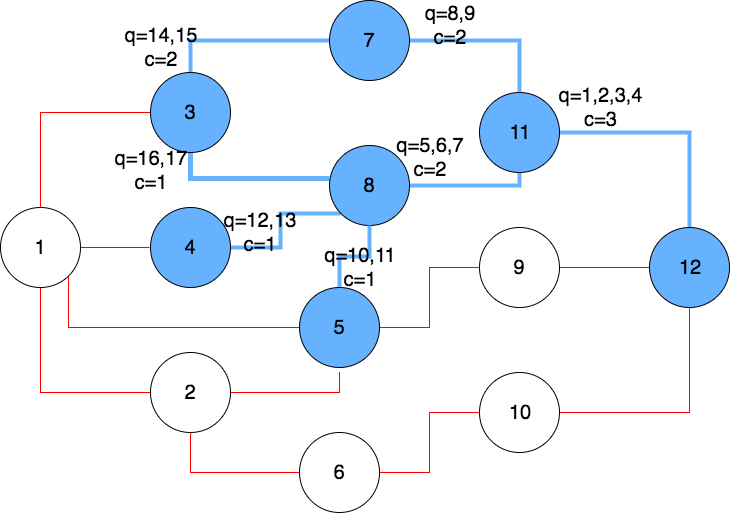}
  \captionof{figure}{Air Traffic Flows of the example system}
  \label{fig:CaseStudy}
\end{minipage}%
\begin{minipage}{.5\textwidth}
  \centering
  \includegraphics[width=1\linewidth]{figures/Numerical11111.png}
  \captionof{figure}{Scenario of complete RDOS attack}
  \label{fig:Sccrds}
\end{minipage}
\begin{minipage}{.5\textwidth}
  \centering
  \includegraphics[width=1\linewidth]{figures/Numerical22222.png}
  \captionof{figure}{Scenario of RST attack }
  \label{fig:Scrst}
\end{minipage}%
\begin{minipage}{.5\textwidth}
  \centering
  \includegraphics[width=1\linewidth]{figures/Numerical33333.png}
  \captionof{figure}{Scenario of SDOS attack}
  \label{fig:Scsds}
\end{minipage}
\end{figure}

\section{Flow Impact Case Studies} 
This section investigates the impact of the proposed attacks on a test system to analyze the queue backlogs and resulting delay. The case study that we use is as follows, at $t=0$, there are 21 airplanes in the system which are in the queues of the sectors. Sectors 3,4, and 5 are the source of aircrafts. The destination of each aircraft is assumed to be sector 12. Fig. \ref{fig:CaseStudy} shows the initial condition of the flow graph, where $c$ represents the capacity of service in each queue and  $q$ represents the number of aircraft in the queue. In the following, we investigate the impacts of different attacks on the flow graph. The routes are as follows:  $route1 = [3,7,11,12]$, $route2 = [3,8,11,12]$, $route3 = [4,8,11,12]$, and $route4 = [5,8,11,12]$.

The impact analysis shows the three scenarios for the RDOS, RST, and SDOS attack. 
\begin{itemize}
\item In the RDOS scenario~\ref{fig:Sccrds}, the attack starts at $t=1$,  targets sector 8, and lasts  one time interval. Therefore, sector 8 cannot serve the aircrafts  at $t=1$. The RST scenario~\ref{fig:Scrst} starts at $t=0$ and targets aircraft number 15. 

\item  In the RST scenario~\ref{fig:Scrst}, the attack starts at $t=0$ to show the impact of the attack on sectors 7 and 8 at $t=1$ and sector 11 at $t=2$ of attack. The attacker creates a non-existing (ghost) aircraft using the ID of airplane 15 between sectors 3 and  8. Simultaneously, airplane number 15 vanishes from the radar screen when it travels from sector 3 to sector 7.

\item In the SDOS scenario~\ref{fig:Scsds}, the attack starts at $t=2$, targets sector 8, and lasts one time interval. Unlike the RDOS attack where the most affected sector is the target sector (sector \#8), in SDOS attack, the most affected sector is the sector 11 where is located after target sector. Therefore, we start the attack at $t=2$ to be able to compare the effect of different attacks on the same sector at the same time. In other words, By starting the RDOS attack at $t=1$ and the SDOS attack at $t=2$, we can see the effect of attacks on sector 11 at the same time.
\end{itemize}

\begin{figure}
\centering
\begin{minipage}{.5\textwidth}
  \centering
  \includegraphics[width=1\linewidth]{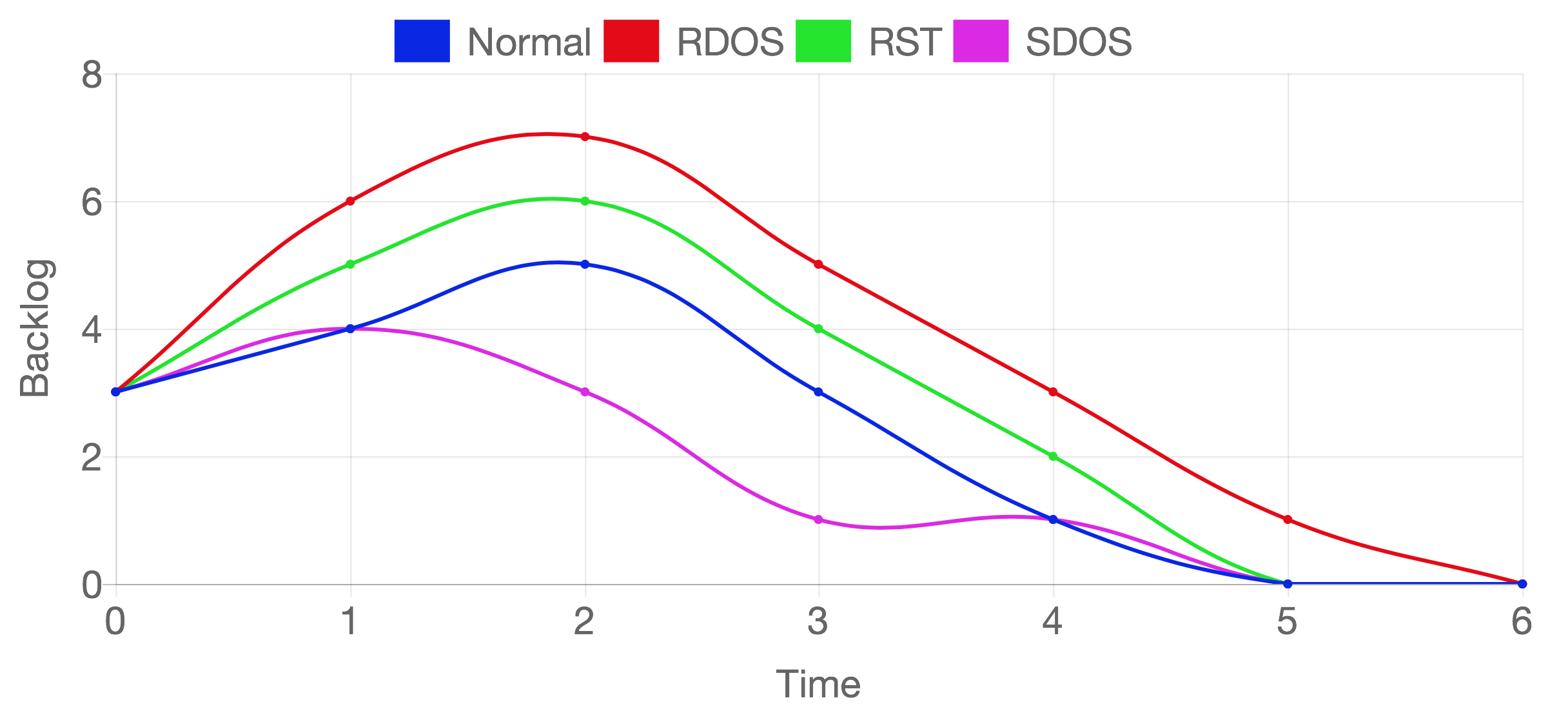}
  \captionof{figure}{$x_{11}^{8}(t)$ in normal and attacks cases} 
  \label{fig:backlog8}
\end{minipage}%
\begin{minipage}{.5\textwidth}
  \centering
  \includegraphics[width=1\linewidth]{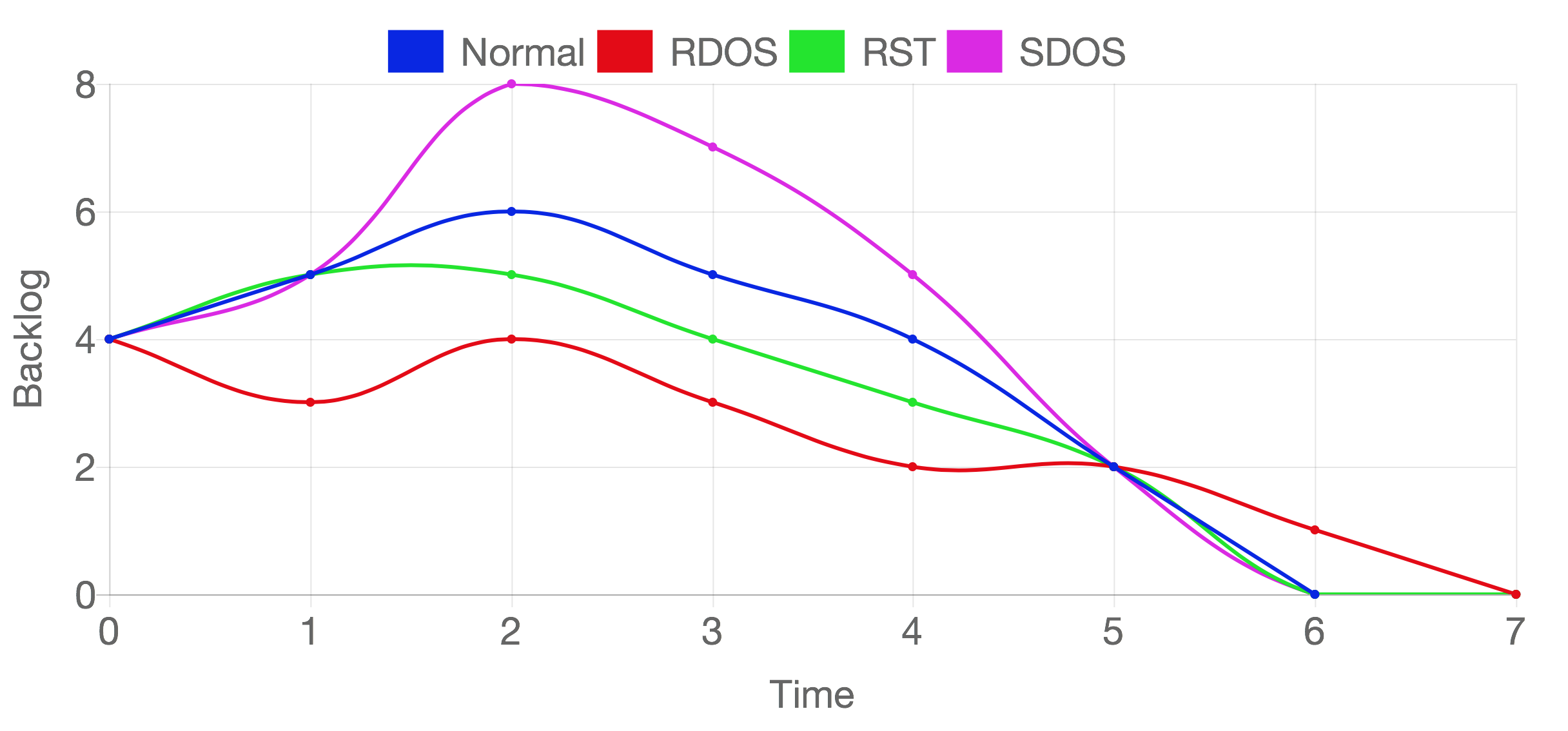}
  \captionof{figure}{$x_{12}^{11}(t)$ in normal and attacks cases}
  \label{fig:backlog11}
\end{minipage}
\end{figure}

For each attack, we extract the backlog $x_{11}^{8}(t)$ for sector 8 and $x_{12}^{11}(t)$ for sector 11 which are located after target of attack. $x_{j}^{i}(t)$ is the backlog of sector $i$ for the outflow which is directed to sector $j$. Fig. \ref{fig:backlog8} and Fig. \ref{fig:backlog11} show $x_{11}^{8}(t)$ and $x_{12}^{11}(t)$ respectively for each of the three attacks. 

In the RDOS attack, since the target sector (sector 8) can not serve the airplanes, the backlog increases. On the other hand, decreasing the inflow of sector 11 leads to decreasing the backlog in this sector during the attack. Therefore, the aircrafts that come to the sector 11 from other inflows (aircraft 15) can be served faster.

In the RST attack, there is a non-existing (ghost) aircraft  in the queue that increases the $x_{11}^{8}(t)$. Since sector 8 schedules a non-existing aircraft, it cannot use all the outflow capacity which is the inflow of sector 11. This issue leads to a decreased backlog of sector 11. The time of arrival of the target aircraft (aircraft 15) increases. In the SDOS attack, all the aircrafts in the queue of the target sector (sector 8) fly to the next sector. It leads to a decreased  backlog of the target sector. On the other hand, since all the aircrafts that left the target sector arrive the sector 11 together, the backlog of sector 11 increases.

\begin{figure*}[h!]
\centering
\makebox[\textwidth]{\includegraphics[width=\paperwidth]
{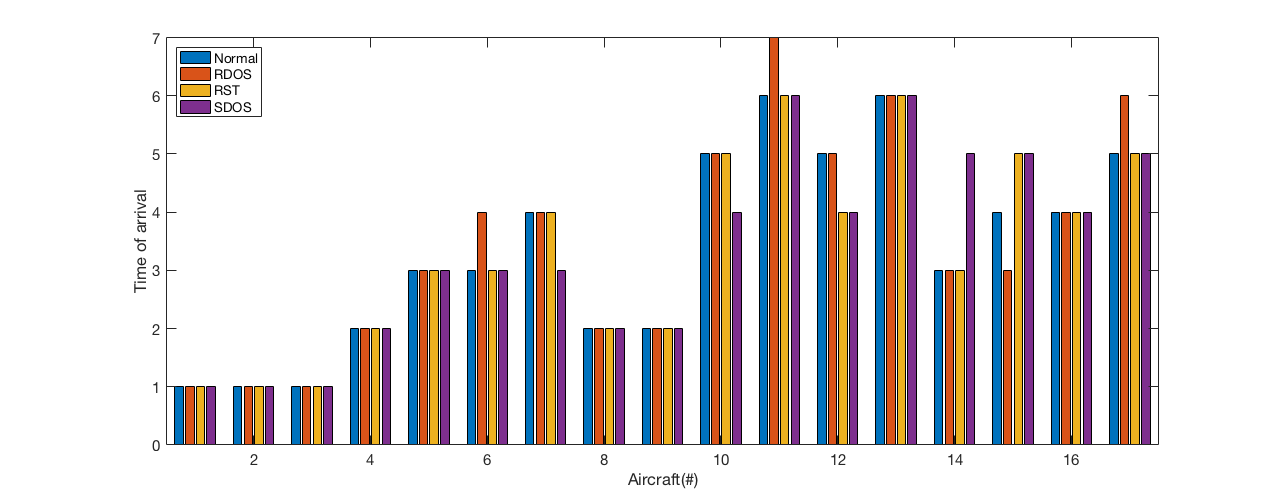}}
\caption{Arrival time of aircrafts in normal and attacks condition }
\label{fig:compareaircrafts}
\end{figure*}

Fig. \ref{fig:compareaircrafts} shows the time of arrival of aircrafts for the normal and attack cases. The results show that the RDOS attack presents the greatest impact to the aircrafts 6, 11 and 17. The RST attack increases the time of arrival of aircraft 15. Finally, the SDOS attack effects on aircrafts 14 and 15.

\section{System-level Vulnerability Analysis}
While the previous section only explored the vulnerability of a small case study, the overall ATC risk needs to evaluate to the entire ATC system, which is a large graph with many sectors. Therefore, we explore  previously identified vulnerability metrics, specifically those proposed in~\cite{royvulnerability}, to determine if they provide a strong indicator of cybersecurity vulnerabilities with ATC. Then, we estimate the vulnerability of the sectors and routes of a given system using the metric.  We present a method to verify the results of the metric to evaluate the impacts of the attacks and compare the results. Finally, we investigate the impacts of attacks on the backlog of the airspace system.

\subsection {Vulnerability Analysis Metric}
In this part, we explain a metric that is proposed in \cite{royvulnerability}. Using the metric, we measure the vulnerability of different sectors and routes. By analyzing the results, we could find which sectors and routes of the model are more vulnerable to different types of attack.

\begin{equation}
V_{T} = \sum_{i=1}^{n-1}\sum_{j=i+1}^{n}\frac{f_{ij}^{\alpha} \mid v_{i} - v_{j} \mid^{\beta}}{\lambda^{c}}
\label{eq:Vtotalmetric} 
\end{equation}

The metric is derived from the flow graph and nominal traffic flow. The procedure for calculating the metric is:
\begin{itemize}
\item \textit{Create Laplacian Matrix (L)}: This matrix is $n \times n$, where $n$ is the number of vertices (sectors) in the flow graph. If there is an edge (route) between vertex $i$ and vertex $j$ the value of $L_{ij}$ is equal to -1. Otherwise, it is 0. The value of diagonal elements ($L_{ii}$) is equal to the value which makes the sum of the row to zero.\newline
\begin{equation}
L_{ij}  =
\left\{
	\begin{array}{ll}
		-1  & \mbox{there is an edge from $i$ to $j$ } \\
		 0 & \mbox{there is no edge from $i$ to $j$ }\\
         - \sum_{j\neq i } L_{ij} & \mbox{$i = j$}
	\end{array}
\right.
\label{eq:laplacian} 
\end{equation}
\item \textit{Calculate eigenvector}: Calculate the eigenvector $v$ which corresponds to the smallest positive eigenvalue.
\item \textit{Vulnerability metric}: Equation \ref{eq:Vmetric} shows the vulnerability metric for the edge between vertex $i$ and vertex $j$. Where $f_{ij}$ is the flow density (number of aircrafts) between sector $i$ and sector $j$. $v_{i}$ and $v_{j}$ are the $i$th and $j$th elements of the eigenvector. the constants $\alpha$ and $\beta$ are positive integers that weight the flows ($f_{ij}$) and eigenvector component differences respectively.
\begin{equation}
    V_{ij} = f_{ij}^{\alpha} \mid v_{i} - v_{j} \mid^{\beta}
\label{eq:Vmetric} 
\end{equation}
\indent If we want to find the total vulnerability metric for the flow graph, we use Equation \ref{eq:Vtotalmetric}. Where $\lambda$ is the eigenvalue and $c$ is a constant that will be tuned based on a formal analysis.
\end{itemize}

\indent We use Equations \ref{eq:Vtotalmetric} and \ref{eq:Vmetric} to determine which of the routes and sectors are more vulnerable. In the rest of this section, we explain how this metrics is used to evaluate the impact of different attacks.

\textbf{1- C-RDOS attack:} In complete RDOS attack, all the outflows are blocked such that no aircraft can pass through outflows. Therefore, in the flow graph model, all routes of the target sector should be removed. This change makes the flow graph disconnected and we can not show the effect of the target sector in the calculation of the vulnerability metric. For the calculating of the metric in this condition, each time we keep one edge of the target node and remove others and calculate the metric using Formula \ref{eq:Vtotalmetric}. The procedure repeats for all edges. If the target node has $m$ edges we do the procedure $m$ times. In this attack, $V_{T}$ is the sum of the calculated values. If the target sector $t$ has $k$ routes and $V_{T_{ti}}$ shows the vulnerability metric when the only available route of target sector ($t$) is $ti$, Equation \ref{eq:CRDOSmetric} shows the total vulnerability metric.
\begin{equation}
    V_{T} = \sum_{i=1}^{k} V_{T_{ti}}
\label{eq:CRDOSmetric} 
\end{equation}

\textbf{2- P-RDOS attack:} In partial RDOS attack, one of the outflows of the target sector is blocked. For calculating the vulnerability metric, first we remove the edge corresponding to the blocked route from the graph. Then calculate $V_{T}$ using Equation \ref{eq:Vtotalmetric}.

\textbf{3- SDOS attack:} In SDOS attack, the outflows of target sector increases and no route is blocked. There is no change in the structure of graph model during this attack and only $f_{ij}$ of the outflows of target sector increases. Therefore, we calculate $V_{T}$ using Equation \ref{eq:Vtotalmetric}.

\subsection{Metric Comparison}
\begin{figure}
\centering
\begin{minipage}{.5\textwidth}
  \centering
  \includegraphics[width=1\linewidth]{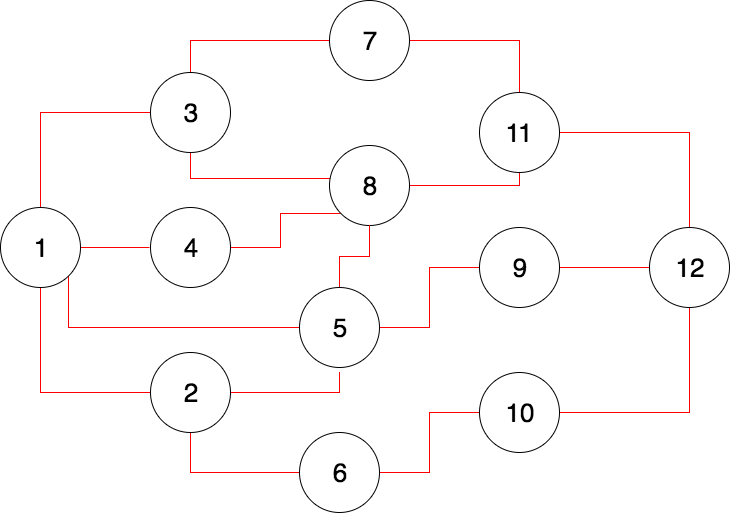}
  \captionof{figure}{Example of air traffic flows}
  \label{fig:structureexp}
\end{minipage}%
\begin{minipage}{.5\textwidth}
  \centering
  \includegraphics[width=1\linewidth]{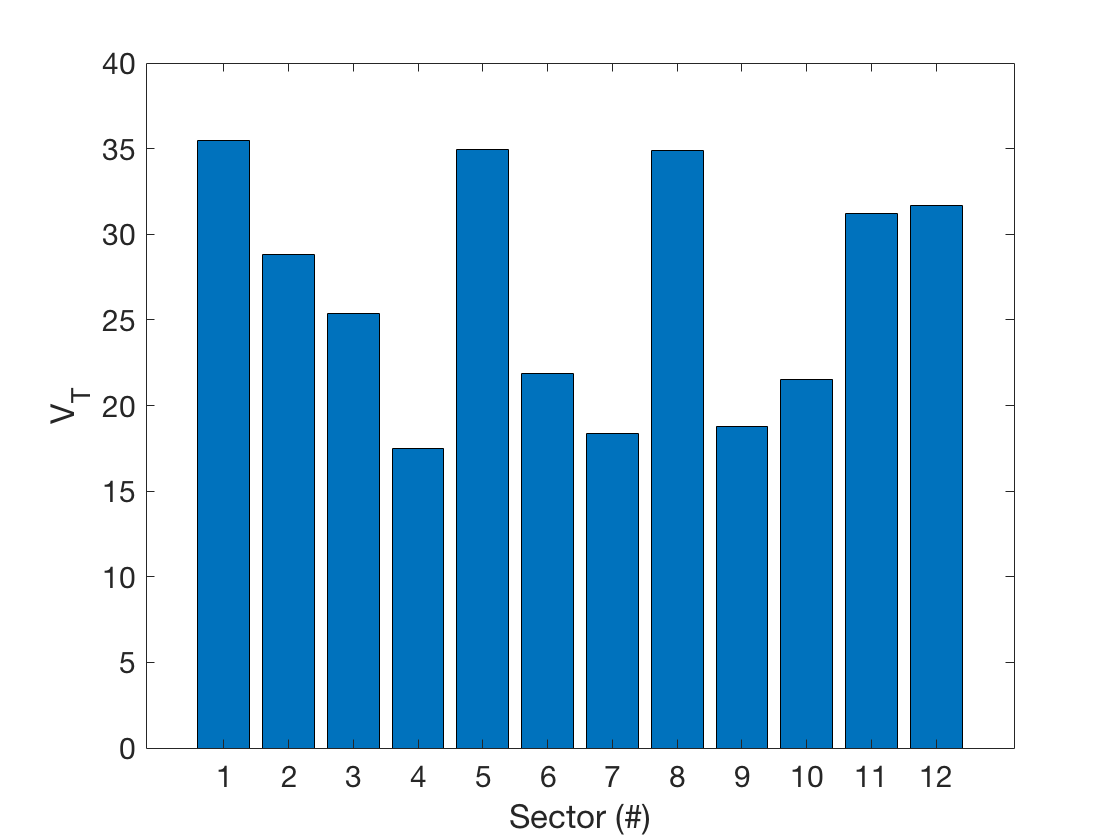}
  \captionof{figure}{$V_{T}$ for each sector which is under complete RDOS attack}
  \label{fig:RDNSresults}
\end{minipage}
\begin{minipage}{.5\textwidth}
  \centering
  \includegraphics[width=1\linewidth]{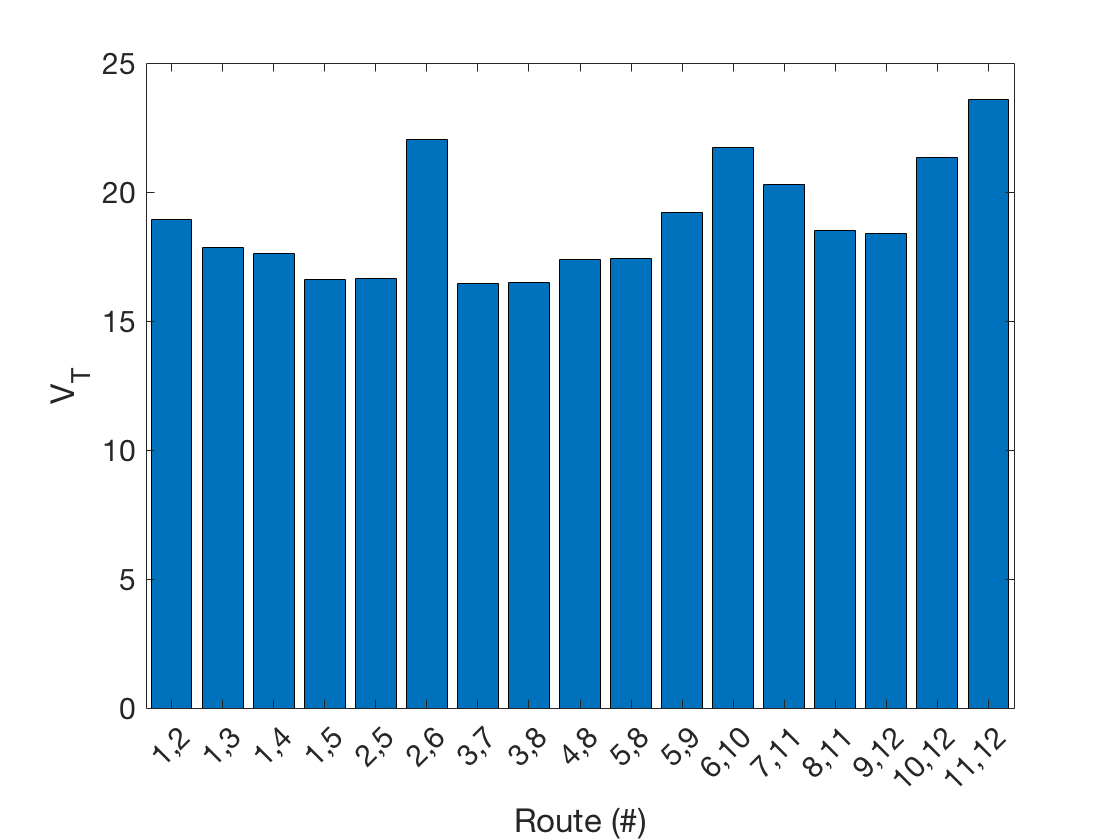}
  \captionof{figure}{$V_{T}$ for each route which is under partial RDOS attack }
  \label{fig:RDNSroute}
\end{minipage}%
\begin{minipage}{.5\textwidth}
  \centering
  \includegraphics[width=1\linewidth]{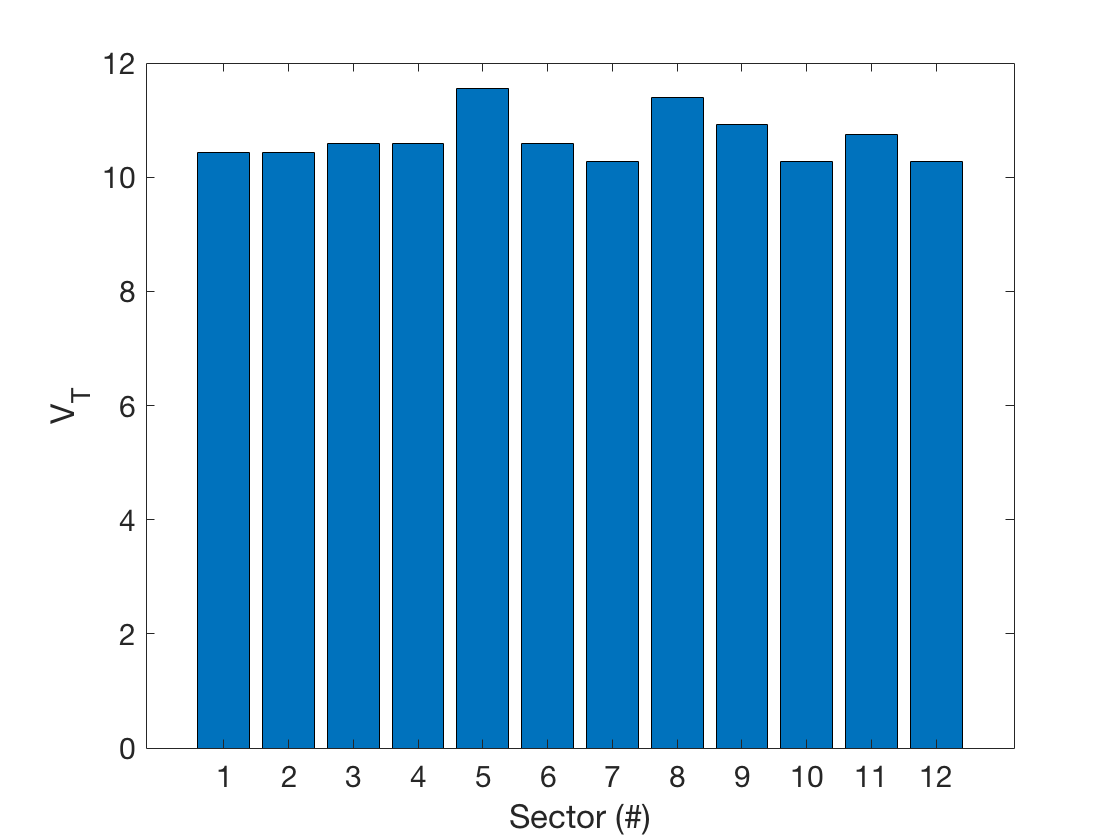}
  \captionof{figure}{$V_{T}$ for each sector which is under SDOS attack}
  \label{fig:SDOSexp}
\end{minipage}
\end{figure}

\indent In this section, we calculate the presented metric for the routes and sectors of the flow graph that is shown in Fig. \ref{fig:structureexp}.

\textit{RDOS attack}: In this experiment, we investigate the impact of the RDOS attack on different sectors. The example of Fig. \ref{fig:structureexp} has 12 sectors. In complete RDOS, each time one of the sectors is the target of attack and then we calculate the metric. By comparing the values, we can find which sectors or routes are more important in the flow graph. In partial RDOS, each time we suppose that one of the routes is the target of attack and calculate the metric. In our experiment, each route is bi-direction. We assume that the value of $f$ for all routes is equal to 2. Fig. \ref{fig:RDNSresults} shows the results of complete RDOS attack and Fig. \ref{fig:RDNSroute} shows the results of partial RDOS attack. The results show that sectors 1, 5 and 8 are more vulnerable than other sectors. Moreover, the routes $11,12$, $2,6$, $6,10$, and $10,12$ are more vulnerable than other routes.

\textit{SDOS attack}: In this experiment, we investigate the impact of the SDOS attack on different sectors. We use the example of Fig. \ref{fig:structureexp} for this experiment. Each time one of the sectors is the target of attack and then calculate the value of $V_{T}$. We suppose that the number of aircrafts in outflows of targeted sector increases by the factor of three. In other words, the normal value of $f$ is equal to 2, and the value of $f$ for outflows of the target sector is equal to 6. Fig. \ref{fig:SDOSexp} shows the results.

\textit{RST attack}: Since the RST attack does not change the structure of flow graph and has a low impact on changing the value of $f$ (only the value of $f$ for two routes is changed), we do not show any results from this analysis.


\subsection{Metric verification for RDOS attack}
In this section, we investigate the structure of the flow graph when a sector is the target of RDOS attack. We present different parameters and find a formula to compute vulnerability of sectors and paths in air traffic flows. Then we compare our results with the values of the metric presented in the previous section ($V_{T}$). 
For calculating vulnerability degree in air traffic flows, we need to know that how many paths are eliminated after shutting down a sector. First we introduce two variables, $lostpath_{k}^{n}$ and $reducepath_{k}^{n}$, as defined below. 

\begin{itemize}
\item $lostpath_{k}^{n}$: If there are $\alpha$ routes with length of $n$ between the sector $i$ and sector $j$, but when sector $k$ is shutdown ($k\neq i$ and $k\neq j$) there is no longer any routes with length of $n$, then $lostpath_{k}^{n}(i,j)=\alpha$. This factor shows when sector $k$ is shutdown, all the routes with length of $n$ between $i$ and $j$ are eliminated. Then, $lostpath_{k}^{n}=\sum_{i \in S} \sum_{j \in S} lostpath_{k}^{n}(i,j)$. Where $S$ is the set of sectors in air traffic flows. We can use same definition to find the number of routes that are eliminated after shutting down a route. In this condition, $k$ is a path that is shutdown.
\begin{equation}
lostpath_{k}^{n}(i,j)  =
\left\{
	\begin{array}{ll}
		\alpha  & \mbox{if all the $a$ routes with length of $n$ between $i$ and $j$ are eliminated} \\
		 0 & \mbox{if at least one route with length of $n$ is remained between $i$ and $j$}
	\end{array}
\right.
\label{eq:laplacian} 
\end{equation}
\item $reducepath_{k}^{n}$: If there are $\alpha$ routes with length of $n$ between the sector $i$ and sector $j$ and when the target sector $k$ is shutdown ($k\neq i$ and $k\neq j$) there remains $\beta$ routes ($0<b<a$) with length of $n$ between $i$ and $j$, $reducepath_{k}^{n}(i,j)=\alpha-\beta$. This factor shows that by  shutting down a sector, some routes with length of $n$ between $i$ and $j$ are eliminated. But $\alpha-\beta$ routes are remained. Then, $reducepath_{k}^{n}=\sum_{i \in S} \sum_{j \in S} reducepath_{k}^{n}(i,j)$. We can use same definition to find the number of routes that are reduced when a route is shutdown. In this condition, $k$ is a path that is shutdown.
\begin{equation}
reducepath_{k}^{n}(i,j)  =
\left\{
	\begin{array}{ll}
		\alpha-\beta  & \mbox{if $\beta$ routes of the all $\alpha$ routes with length of $n$ between $i$ and $j$ are remained} \\
		 0 & \mbox{if all the routes or no routes with length of $n$ are eliminated between $i$ and $j$}
	\end{array}
\right.
\label{eq:laplacian} 
\end{equation}
\item $defaultpath^{n}$: The number of routes with length of $n$ in air traffic flow graph.
\end{itemize}
Using presented parameters, we can define a measure for vulnerability of air traffic flows.

\begin{equation}
V_{k}=\sum_{i=1}^{max(n)}(max(n)-i+1)\frac{[\lambda( lostpath_{k}^{i})+(1-\lambda)( reducepath_{k}^{i})]}{defaultpath^{i}}
\label{eq:mymetric} 
\end{equation}

Where $V_{k}$ is the vulnerability of the air traffic flows when sector $k$ (or route $k$) is shutdown. The term $(max(n)-i+1)$ defines the weight of the routes based on their length. The constants $\lambda$ is $0.5<\lambda<1$ and weight the sensitivity of lost paths respect to reduced paths. We use the example of Fig. \ref{fig:structureexp} and calculate $V_{k}$ for each sector and route. We use $\lambda = 0.75$ in our calculations.
Table \ref{tab:comparemetric} shows the comparison between our metric ($V_{k}$) and the metric presented in \cite{royvulnerability} ($V_{T}$). The rank column shows which sectors are more vulnerable.  The results show that the $V_{K}$ metric recognizes the sector \#5 and the sector \#1. The $V_{T}$ metric identifies same sectors as the most vulnerable sectors but in the different order (sector \#1 as the first rank and sector \#5 as the second rank). The main reason is that they have more routes compared with other sectors. Moreover, some of the paths that pass through them are not replaceable by other paths.  Both metrics rank the sectors \#8 and \#12 as the third and fourth vulnerable sectors. Sector \#8 has four routes (same as sectors \#1 and \#5); however, it is less vulnerable than sector \#1 and \#5. The main reason is that the paths of sector \#8 are less critical than the paths of sectors \#1 and \#5. For example, by shutting down the sector \#5 we lose a path with the length of three between sectors \#1 and \#12 that is not replaceable by other paths. But, by shutting down the sector \#8 all the paths that pass through this sector could be replaced by other paths. For example, the path between sector \#1 and \#11 which goes through sector \#8 could be replaced by the path that passes through sector \#3 and \#7 and has the same length. By comparing the results, we can find that the difference of the rank in both metrics is at most 1.

\begin{table}[h!]
\begin{footnotesize}
\begin{center}
 \begin{tabular}{|c|c|c|c|c|c|}\hline
 Sector & \multicolumn{2}{|c|}{$V_{K}$}  &\multicolumn{2}{|c|} {$V_{T}$} & Difference \\ \hline
     	 &Rank &Value & Rank & Value & \\ \hline
        1 & 2&15.68 & 1& 35.49 & 1 \\ \hline
        2 & 5&12.1 & 6&28.82 & 1 \\ \hline
        3 & 7&8.56 & 7&25.38 & 0 \\ \hline
        4 & 12&3.36 & 12&17.5 & 0 \\ \hline
        5 & 1&20.48 & 2&34.95 & 1 \\ \hline
        6 & 8&5.87 & 8&21.88 & 0 \\ \hline
        7 & 10&5.56 & 11&18.38 & 1 \\ \hline
        8 & 3&14.92 & 3&34.9 & 0 \\ \hline
        9 & 11&5.25 & 10&18.79 & 1 \\ \hline
        10 & 9&5.69 & 9&21.53 & 0 \\ \hline
        11 & 6&11.71 & 5&31.2 & 1 \\ \hline
        12 & 4&12.45 & 4&31.65 & 0 \\ \hline
      \end{tabular}
\caption{Compare metrics for Complete RDOS attack}
\label{tab:comparemetric}
\end{center}
\end{footnotesize}
\end{table}

\subsection{Long-term Impacts on Air Traffic Flow}
\begin{figure}
\centering
\begin{minipage}{.5\textwidth}
  \centering
  \includegraphics[width=.95\linewidth]{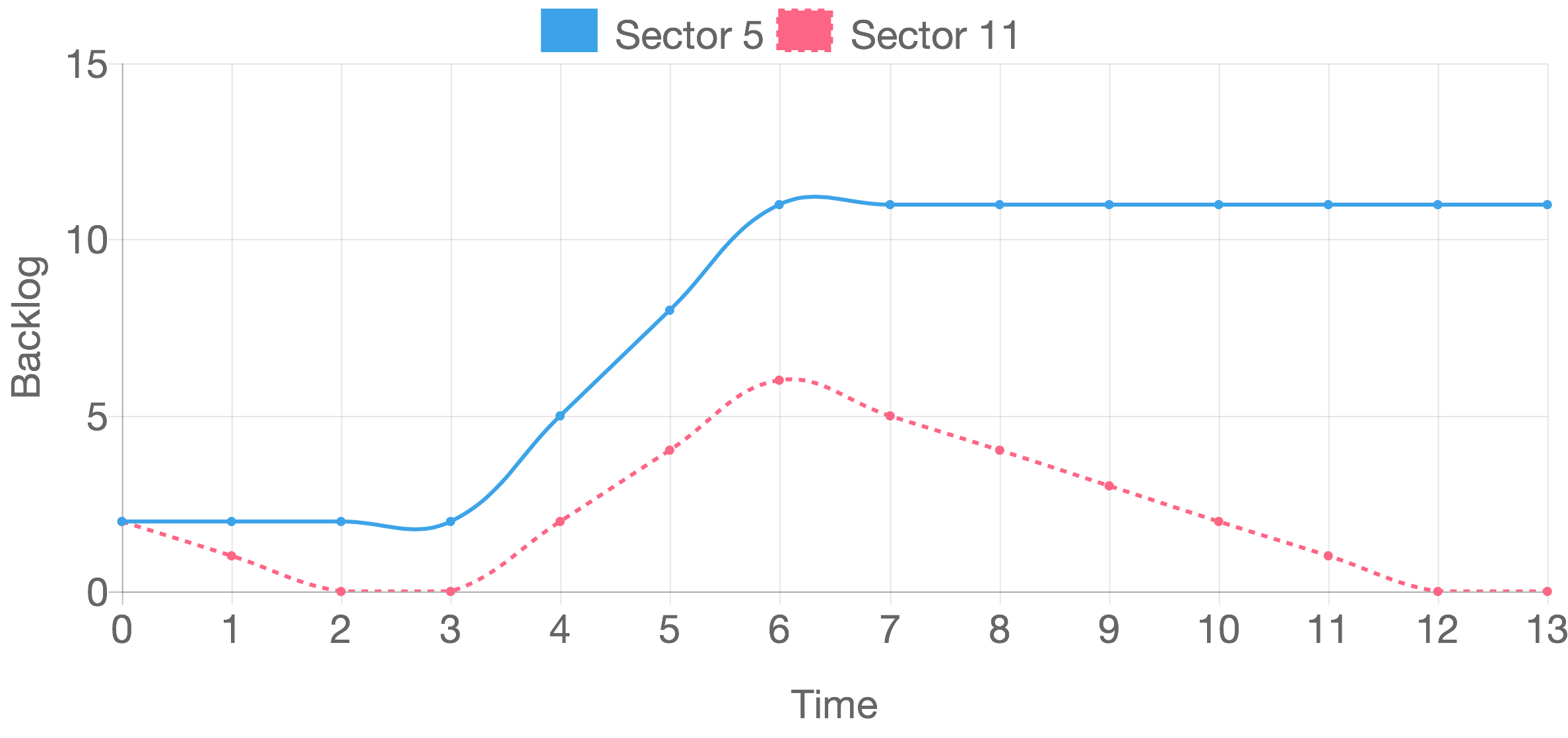}
  \captionof{figure}{Backlog when for the outflow of sector 5 $U_i(t) = C_i(t)$ and for the outflow of sector 11 $U_i(t) < C_i(t)$}
  \label{fig:case1}
\end{minipage}%
\begin{minipage}{.5\textwidth}
  \centering
  \includegraphics[width=.95\linewidth]{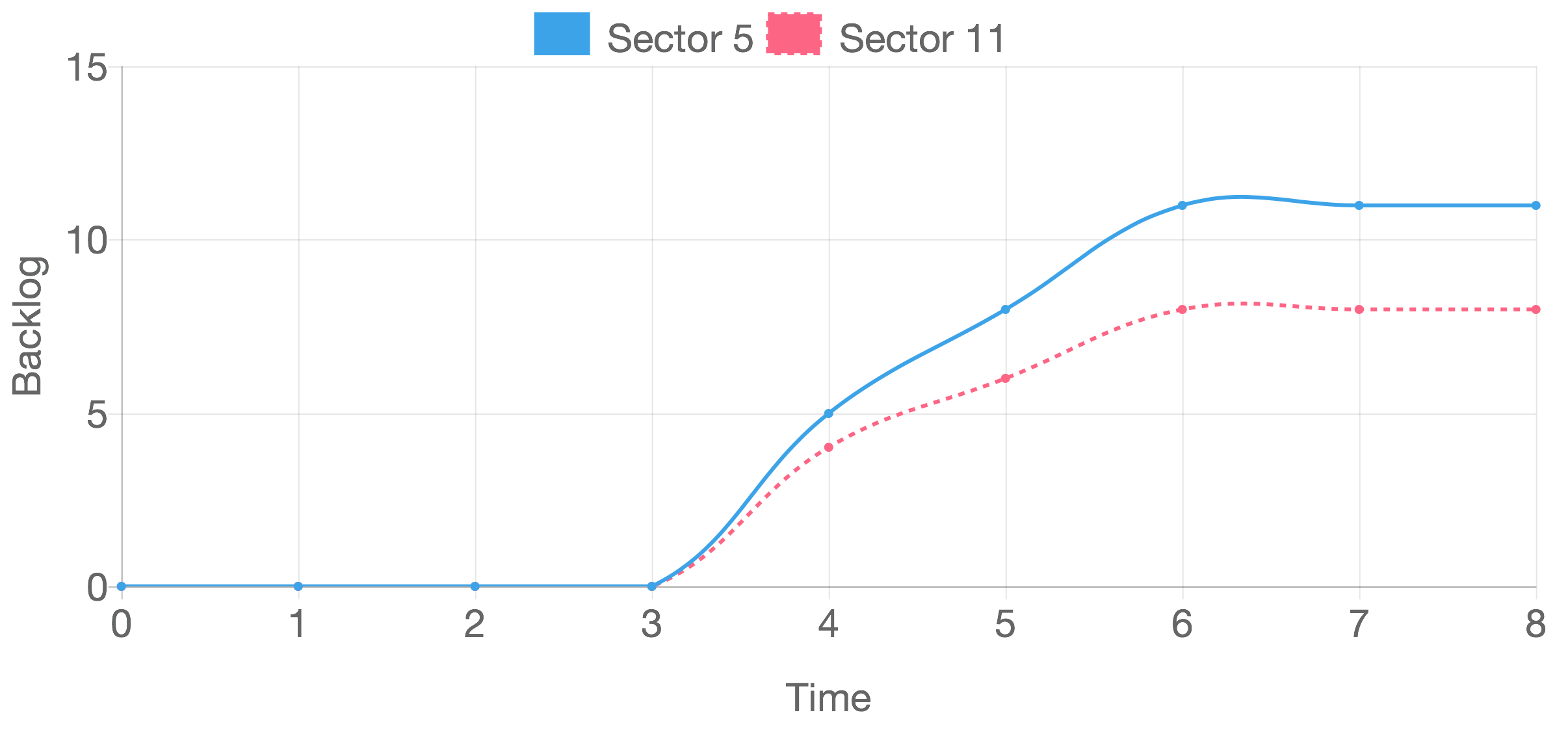}
  \captionof{figure}{Backlog when for the outflow of sector 5 $U_i(t) = C_i(t)=3$ and for the outflow of sector 11 $U_i(t) = C_i(t)=2$}
  \label{fig:case1-2}
\end{minipage}
\end{figure}

\indent \indent The factors such as the capacity of outflow ($C_{i}$) and the size of the input of queue ($U_{i}$) change the impact of an attack in the long-term. During the attack the value of the backlog increases and after the attack it does not be changed, but it is greater than the value before the attack. In this part, we investigate the impacts of attacks in different cases for the complete RDOS. In this attack, we consider two different scenarios to compare the results. In one scenario $U_i(t) < C_i(t)$ and in the second scenario $U_i(t) = C_i(t)$.  We use the flow graph of Fig. \ref{fig:structureexp} for our examples. The properties of attack are as follow.
\begin{itemize}
\item The attack starts at $t=3$ and lasts until $t=6$. For the first case the routes of sector 5 are the target of attack and for the second case, the routes of sector 11 are the target of attack. For sector 5 we focus on the outflow to sector 9 and for sector 11 we focus on the outflow to sector 12.
\item $X_{9}^{5}(0)=X_{12}^{11}(0)=2$. The initial backlog of the queue for both cases is 2.
\item $C_{9}^{5}(t)=C_{12}^{11}(t)=3$. In each time interval, three aircrafts are served.
\item In the first case $U_{9}^{5}(t)=3$ and for the second case $U_{12}^{11}(t)=2$. It is the number of aircrafts which come to the queue in each time interval.
\end{itemize}
Fig. \ref{fig:case1} shows how the attack affects the backlog. The result shows that if $U_i(t) < C_i(t)$, the sector returns to normal condition after some time intervals. But if $U_i(t) = C_i(t)$, effects of the attack remains.
\indent In another experiment, we investigate the rate of backlog increasing. We keep all the properties from the previous experiment and only change $C_{12}^{11}=2$ for the second case. In both cases $U_i(t) = C_i(t)$ but has different values. Fig. \ref{fig:case1-2} shows how the attack affects the backlog. The results of two experiments show that if $U_i(t) = C_i(t)$, the backlog increases by the rate of $U_i(t)$ during the attack. If $U_i(t)<C_i(t)$, first the backlog decreases with rate of $U_i(t) - C_i(t)$ and during the attack it increases with rate of $U_i(t)$.

\section{Discussion and Potential Mitigations}

While the previous sections introduced a combination of analytical attack scenarios and vulnerability metrics to evaluate the impact of attacks to ATC, this section will provide further discussion identifying the key factors contributing to the risk of these attacks based on the significance of the route manipulations. During the complete RDOS attack, all the outflows are blocked and all of  the inflows should wait in the queues and increase the backlog of the queues. Therefore, the rate of inflows during the attack has a significant role in the impact of attacks. If the rate of inflows is $m$ aircrafts per time interval and the sector has $n$ outflows, The backlog of queues increases by the rate of $m/n$ aircrafts per time interval averagely. The rate for each outflow $i$ is $U_i$ which is the input of the queue of outflow. The value of $C$ is an important factor to decrease the impacts of the attack when it is finished. If there are $n$ outflows and $\sum_{i}^{n} C_i = c$, the aircrafts are served by the rate of $c/n$ per time interval averagely. The rate for each outflow is $C_i$. For partial RDOS attack, the impact is only on one outflow. 

The target of a RST attack is an aircraft. Therefore, there is not any blockage in the inflows or outflows. As the attacker is only injecting a fake route for the target aircrafts, the backlog of one of the outflows increases one unit.

For SDOS attack, as a result of removing the queue management, the outflows increase. If the attack happens, all the aircraft in the queue and all of the new aircrafts pass through outflows. after that, there is no aircraft in the queue and all the new aircrafts pass to outflows. The rate of increasing the outflow in the first time interval of attack is $b_{i}+U_{i}$ and after that is $U_{i}$. This large amount of outflows impacts on the inflows of the next sectors and leads to increasing the backlog of them.

\indent  The RDOS attack increases the backlog of the target sector. RST attack changes the backlog of original destination and new destination which is injected by the attacker. In the SDOS attack, the backlog increases for the sectors that have a route from the targeted sector. Table \ref{table:attackimpact} shows a summary of attacks. We know the value of $U_i(t)$ effects the backlog in different attacks. Fig. \ref{fig:inout} shows the effects of $U_i(t)$ to the sectors. The blue line shows the effect when $U_i(t) < C_i$. In other words, the number of aircrafts that enter the queue is smaller than the number of aircrafts get served. In this condition, the backlog decreases to zero in some time intervals after the attack. During the attack the value of backlog increases. When attack ends, since $U_i(t) < C_i$ the value of backlog decreases. The green dash line shows the effect of attack when $U_i(t) = C_i$. When the attack happens, the backlog increases. After the attack, since $U_i(t) = C_i$ the value of backlog remains constant, as the effect of attack continues. The dotted red line shows the condition that $U_i(t) > C_i$. In this condition, the value of the backlog increases before the attack. During the attack, the value of backlog continues to increase with the larger rate. When the attack finishes, the rate of increasing returns to the normal value.


\begin{table}[ht]
\begin{minipage}[b]{0.5\linewidth}
\begin{footnotesize}
\centering
\begin{tabular}{ | c | c | c | c |}
    \hline
    Attacks &RDOS & RST & SDOS \\ \hline
    Effected& inflow & outflow & outflow \\ 
    by& &(minor) & \\ \hline
    Backlog & & & \\
    increase& $U_i$ & 1 & $b_{i}+U_{i}$  \\
    rate & & & \\ \hline
    Impacts & single  & 2  & multi  \\ 
    on & sector & sectors & sectors \\ \hline
   \end{tabular}
    \caption{Attacks impact summary}
    \label{table:attackimpact}
\end{footnotesize}
\end{minipage}\hfill
\begin{minipage}[b]{0.5\linewidth}
\centering
\includegraphics[width=1\linewidth]{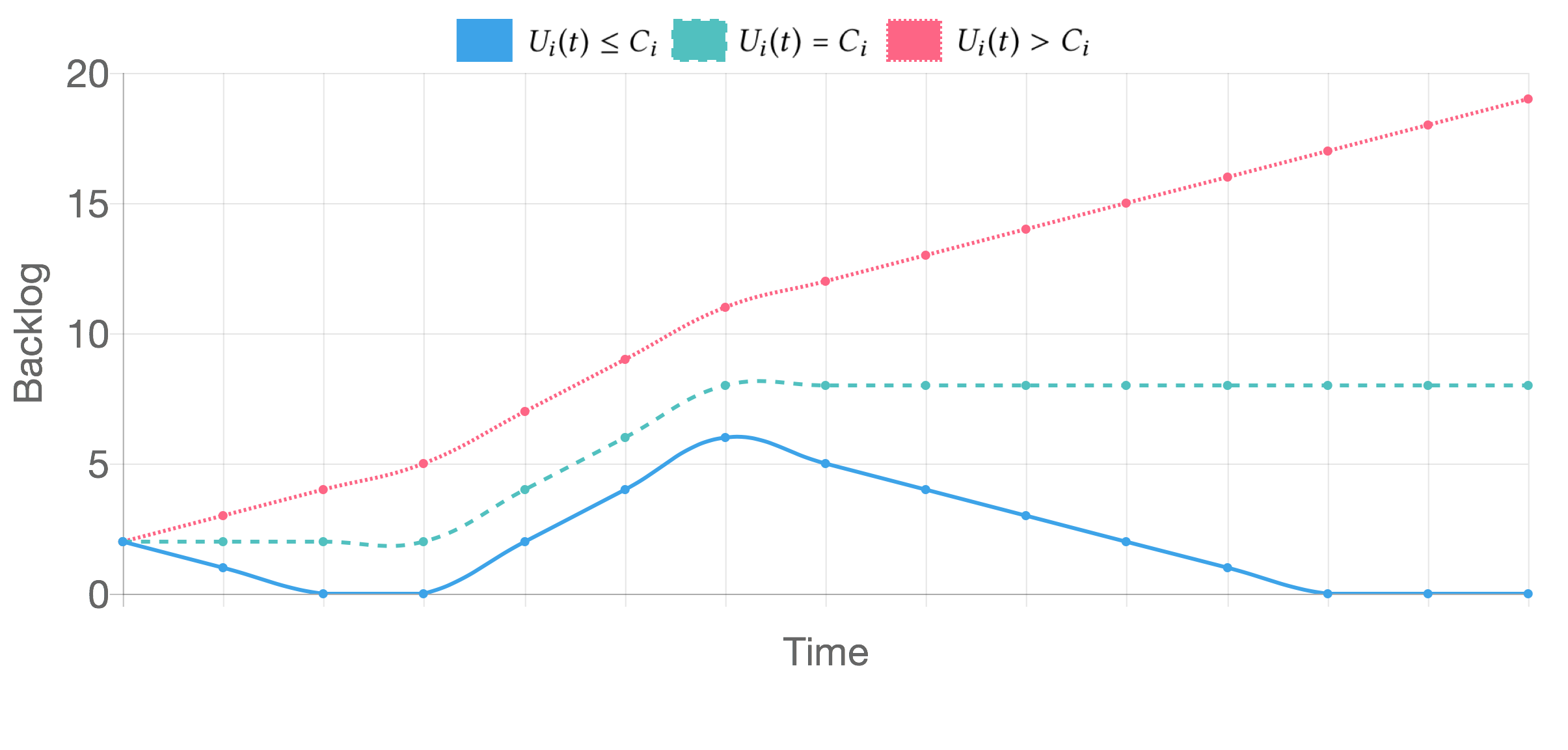}
\captionof{figure}{How the value of $U_i(t)$ effects on backlog}
\label{fig:inout}
\end{minipage}
\end{table}
Table \ref{tab:comparetable} shows a conclusion of the proposed attacks, including their goal, difficulty, and attack vector.
\begin{table}[h!]
\begin{center}
 \begin{tabular}{ |c|c|c|c|c| }
\hline
Attack&Impact&Difficulty&Attack vector&Goal\\
\hline
RDOS&High&Medium&Message Injecting& Shutdown route or sector\\
\hline
RST&Low&High&Message Jamming and injecting& creating a fake route\\
\hline
SDOS&Medium-High&Low&Jamming&Bypass queuing\\
\hline
\end{tabular}
\caption{Attacks conclusion}
\label{tab:comparetable}
\end{center}
\end{table}

\subsection{Attack Mitigation Techniques}
There are several countermeasures such as secure location verification and secure broadcast authentication methods that help to increase the security of ADS-B system and avoid these attacks \cite{strohmeier2015security}, however, these methods have some drawbacks. Secure location verification methods usually need expensive systems which would not be cost-effective for ADS-B. The secure authentication methods face the problems such as small length of ADS-B message that does not have enough space for more headers (e.g, message authentication code). Furthermore, the large number of aircrafts and they're wide geographic dispersion across multiple countries and their ATC domains  makes the key management and distribution more difficult \cite{strohmeier2015security}.

In lack of the attack mitigation techniques, rerouting the blocked aircraft could help to decrease the backlog in RDOS attack.approaches may exist to mitigate the impacts of these attacks. In SDOS attack, Increasing the value of $c$ for a short period in the sectors which are located after the target sector leads to decreasing the backlog with the more significant rate.

\section{Conclusions}
In this paper we explained three different attack scenarios (RST, RDOS, SDOS) to air traffic control systems, develop a formal DQN model to evaluate their impact on a simplified model. The RDOS attack is launched by injecting the ADS-B message of non-existing aircrafts into air traffic system. While the controllers can not distinguish real and injected aircrafts, they don't serve the aircrafts and the routes of the sector are shutdown. As a result of this attack, the outflows are blocked and the aircrafts have to wait to be served. The RST attack blocks ADS-B message from a valid aircraft and injects manipulated ADS-B message for the airplane. This attack affects on two outflows of the sector and makes changes in queue management of the sector controller. The SDOS attack is lunched by jamming the messages near the sector controller. As a result of this attack, aircrafts are unable to receive the commands of the sector about queue management and pass the sector without waiting in queue. Based on this analysis we demonstrate that the RDOS attack provides the greatest impact to the ATC routes and will likely introduce the greatest delays, while the SDOS and RST attack maintain more confined impact to specific planes and routes.

\bibliographystyle{ACM-Reference-Format}
\bibliography{sample-bibliography}

\end{document}